\documentclass[twocolumn]{aastex631}

\usepackage{natbib}
\usepackage{epsfig}
\usepackage{graphicx}
\usepackage{subfigure}
\usepackage{float}
\usepackage{amsmath}
\usepackage{color}
\usepackage{amssymb}
\usepackage{apjfonts}

\newcommand{\PMO}{Purple Mountain Observatory, Chinese Academy of Sciences, Nanjing 210023, China}
\newcommand{\USTC}{School of Astronomy and Space Sciences, University of Science and Technology of China, Hefei 230026, China}
\newcommand{\HNU}{School of Physics and Electronics, Hunan Normal University, Changsha 410081, China}

\shortauthors{Wang \& Wei}

\graphicspath{{./}{figures/}}

\usepackage{amsmath}
\begin{document}

\title{An 8.0\% Determination of the Baryon Fraction in the Intergalactic Medium from Localized Fast Radio Bursts}


\correspondingauthor{Jun-Jie Wei}
\email{jjwei@pmo.ac.cn}

\author[0000-0003-3635-5375]{Bao Wang}
\affiliation{\PMO}
\affiliation{\USTC}
\affiliation{\HNU}

\author[0000-0003-0162-2488]{Jun-Jie Wei}
\affiliation{\PMO}
\affiliation{\USTC}

\begin{abstract}
The dispersion measure (DM)--redshift relation of fast radio bursts (FRBs) has been proposed as
a potential new tool for probing intergalactic medium (IGM) and for studying cosmology. However,
the poor knowledge of the baryon fraction in the IGM ($f_{\mathrm{IGM}}$) and its degeneracy
with cosmological parameters impose restrictions on the cosmological applications of FRBs.
Furthermore, DMs contributed by the IGM ($\mathrm{DM_{IGM}}$) and host galaxy ($\mathrm{DM_{host}}$),
important cosmological quantities, cannot be exactly extracted from observations, which
would bring uncontrolled systematic uncertainties in FRB cosmology. In this work, we use seventeen
localized FRBs to constrain $f_{\mathrm{IGM}}$ and its possible redshift evolution. Other
cosmological probes such as type Ia supernovae, baryon acoustic oscillations, and cosmic microwave
background radiation are combined to break parameter degeneracy. Taking into account the probability
distributions of $\mathrm{DM_{IGM}}$ and $\mathrm{DM_{host}}$ derived from the the IllustrisTNG simulation,
we obtain a robust measurement of $f_{\mathrm{IGM}}=0.927\pm0.075$, representing a precision of 8.0\%.
We find that there is no strong evidence for the redshift dependence of $f_{\mathrm{IGM}}$ at the
current observational data level. The rapid progress in localizing FRBs will significantly improve
the constraints on $f_{\mathrm{IGM}}$.
\end{abstract}


\section{INTRODUCTION}
\label{Sec:Intro}
Fast radio bursts (FRBs) are a class of brief ($\sim$ms) and intense ($\sim$Jy) radio transients
with large dispersion measures (DMs), well in excess of the expected contributions from the Milky Way
\citep{Lorimer2007,Thornton2013,Petroff2016,Platts2019,Xiao2021,Zhang2022}. Owing to their anomalously
high DMs, FRBs are believed to be of extragalactic or even cosmological origin. To date,
more than 600 FRBs have been detected, and over two dozen of them have been reported to repeat
\citep{2021ApJS..257...59C}. There are more than 20 FRBs with definite host galaxies and redshift measurements.
These observations suggest that FRBs are promising tools for studying cosmology. Some proposals
include using localized FRBs to measure the baryon number density of the universe
\citep{Deng2014,McQuinn2014,Macquart2020, Yang2022}, constrain the dark energy equation of state
\citep{Gao2014, Zhou2014, Walters2018, 2018ApJ...860L...7W,Zhao2020,2022JCAP...02..006Q},
constrain cosmic reionization history \citep{Deng2014,Zheng2014,Hashimoto2021}, measure
cosmological distance \citep{Yu2017,Kumar2019}, or measure the Hubble constant
\citep{Hagstotz2022, Wu2022}; using strongly lensed FRBs to probe the nature of compact dark matter \citep{2016PhRvL.117i1301M,2018A&A...614A..50W}, or measure the Hubble constant and cosmic curvature
\citep{Li2018}.

Another cosmological puzzle is the baryon distribution of the universe.
While it is widely believed that more than three-quarters of the baryonic content of the universe
resides in the diffuse intergalactic medium (IGM), with only a small fraction in galaxies and
galaxy clusters \citep{1998ApJ...503..518F,2006ApJ...650..560C}, gaining direct observational
evidence of the baryon distribution is challenging. There have been many studies of detecting
the baryon fraction in the IGM, $f_{\mathrm{IGM}}$, through numerical simulations  \citep{1999ApJ...514....1C,2006ApJ...650..560C,Meiksin2009} or observations
\citep{1998ApJ...503..518F,2004ApJ...616..643F,Shull2012,2016PhRvL.117e1301H,2018PhRvD..98j3518M}.
For instance, \cite{Meiksin2009} performed numerical simulations and suggested that $\sim90$\%
of the baryons produced by the Big Bang are contained within the IGM at redshifts of $z\ge1.5$ (i.e.,
$f_{\mathrm{IGM}}\approx0.9$). It was observed that $18\pm4$\% of the baryons
exists in galaxies, circumgalactic medium, intercluster medium, and cold neutral gas at $z\le0.4$,
or equivalently $f_{\mathrm{IGM}}\approx0.82$ \citep{Shull2012}. There is an ongoing debate
about the value of $f_{\mathrm{IGM}}$.

The observed DMs of cosmological FRBs are mainly contributed by the IGM. Since the DM
contributed by the IGM ($\mathrm{DM_{IGM}}$) carries important information on the location
of baryons in the late universe, one may combine the $\mathrm{DM_{IGM}}$ and $z$ information
of FRBs to constrain the IGM baryon fraction $f_{\mathrm{IGM}}$. Indeed, a number of methods
have been proposed to estimate $f_{\mathrm{IGM}}$ by utilizing the DM($z$) data of FRBs
\citep{Li2019,Li2020,Wei2019,Walters2019,Qiang2020,Dai2021,Lin2022,Lemos2022}. However,
one issue that restricts such studies is the strong degeneracy between cosmological parameters
and $f_{\mathrm{IGM}}$. It is hard to determine $f_{\mathrm{IGM}}$ directly only relying on
FRB data. Moreover, there is another thorny issue that DMs contributed by FRB host galaxies
($\mathrm{DM_{host}}$) and the inhomogeneities in the IGM ($\mathrm{DM_{IGM}}$) cannot be exactly
extracted from observations \citep{Macquart2020}. Previous studies assumed fixed values for
them, which would bring uncontrollable systematic uncertainties in the analysis. A more plausible
approach is to treat them as probability distributions derived from cosmological simulations \citep{Jaroszynski2019,Macquart2020,Zhang2020,Zhang2021,Wu2022}.

In this paper, we present a high-precision measurement of the baryon fraction in the IGM
with seventeen localized FRBs through the $\mathrm{DM_{IGM}}$-$z$ relation. In order to break
the degeneracy between cosmological parameters and $f_{\mathrm{IGM}}$, we combine FRB data
with current constraints from type Ia supernovae (SNe Ia), baryon acoustic oscillations (BAOs),
and cosmic microwave background (CMB) radiation. In our $f_{\mathrm{IGM}}$ estimation,
the reasonable probability distributions of $\mathrm{DM_{host}}$ and $\mathrm{DM_{IGM}}$
derived from the IllustrisTNG simulation \citep{Zhang2020,Zhang2021} are adopted to reduce
the systematic errors. Additionally, to explore the possible evolution of $f_{\mathrm{IGM}}$
with redshift, we also consider two different parametric models, namely a constant model
and a time-dependent model.

The rest of our paper is organized as follows. In Section \ref{Sec:Dispersion}, we review
the FRB DM measurements and the probability distributions of $\mathrm{DM_{host}}$ and $\mathrm{DM_{IGM}}$
derived from the IllustrisTNG simulation. In Section \ref{Sec:probes}, we give an introduction of
the compilations of the three other cosmological probes, including SNe Ia, BAO, and CMB. Monte Carlo
Markov Chain (MCMC) parameter inference results are presented in Section \ref{Sec:Results}. Finally,
conclusions are summarized in Section \ref{Sec:Conclusion}.

\section{FRB Dispersion Measures}
\label{Sec:Dispersion}
The precise localization of FRBs to their host galaxies provides an ensemble of DM and $z$ measurements.
The DM measurement represents the integrated column density of free electrons along the line of sight.
For an extragalactic FRB, its observed DM can be separated into the following components:
\begin{equation}
	\mathrm{DM_{obs}}(z)=\mathrm{DM_{ISM}^{MW}}+\mathrm{DM_{halo}^{MW}}+\mathrm{DM_{IGM}}(z)
	+\frac{\mathrm{DM_{host}}}{1+z}\;,
\end{equation}	
where $\mathrm{DM_{ISM}^{MW}}$, $\mathrm{DM_{halo}^{MW}}$, $\mathrm{D M_{IGM}}$, and $\mathrm{DM_{host}}$
represent the DM contributions from the Milky Way interstellar medium (ISM), the Milky Way halo,
the IGM, and the FRB host galaxy, respectively. The $(1+z)$ factor converts the local $\mathrm{DM_{host}}$
to the observed one \citep{Deng2014}.
Because of the inhomogeneity of the free electron distribution in the IGM, two sources at the same
redshift but in different sightlines will likely have significant differences in the measured value
of $\mathrm{DM_{IGM}}$. Adopting the flat $\Lambda$CDM cosmological model, the average value of
$\mathrm{DM_{IGM}}$ can be estimated as \citep{Deng2014}
\begin{equation}\label{DMIGM}
	\left\langle\mathrm{DM}_{\mathrm{IGM}}\right\rangle(z)=\frac{3 c \Omega_{b} H_{0}}{8 \pi G m_{p}} \int_{0}^{z} \frac{(1+z^{\prime}) f_{\mathrm{IGM}}(z^{\prime}) \chi_{e}(z^{\prime})}{\sqrt{\Omega_{m}\left( 1+z^{\prime}\right)^3+1-\Omega_{m}}} d z^{\prime}\;,
\end{equation}
where $m_p$ is the proton mass, $H_0$ is the Hubble constant, $f_{\mathrm{IGM}}(z)$ is the
baryon fraction in the IGM, and $\Omega_{b}$ and $\Omega_{m}$ are the present-day baryon and
matter density parameters. The free electron number fraction per baryon is
$\chi_{e}(z)=\frac{3}{4} \chi_{e,\mathrm{H}}(z)+\frac{1}{8}\chi_{e, \mathrm{He}}(z)$, where
$\chi_{e,\mathrm{H}}(z)$ and $\chi_{e,\mathrm{He}}(z)$ are the ionization fractions of hydrogen
and helium, respectively. Both hydrogen and helium are completely ionized at redshifts $z<3$
\citep{Meiksin2009, Becker2011}, allowing one to set $\chi_{e,\mathrm{H}}= \chi_{e,\mathrm{He}}=1$,
which gives $\chi_{e}=7/8$.

The $\mathrm{DM_{IGM}}$ value of a well-localized FRB can be extracted using $\mathrm{DM_{IGM}}=\mathrm{DM_{obs}}-\mathrm{DM_{ISM}^{MW}}-\mathrm{DM_{halo}^{MW}}-\mathrm{DM_{host}}/(1+z)$.
Here the $\mathrm{DM_{ISM}^{MW}}$ term can be well estimated from the NE2001 model of
the ISM free electron distribution \citep{Cordes2002}. The $\mathrm{DM_{halo}^{MW}}$ term is not well constrained, but is
expected to contribute $50\sim 80$ $\mathrm{pc\;cm^{-3}}$ \citep{Prochaska2019a}. Hereafter we assume
that the probability distribution of $\mathrm{DM_{halo}^{MW}}$ can be described by a Gaussian distribution
with mean $\mu_{\mathrm {halo}}=65$ $\mathrm{pc\;cm^{-3}}$ and standard deviation
$\sigma_{\mathrm {halo}}=15$ $\mathrm{pc\;cm^{-3}}$ \citep{Wu2022}:
\begin{equation}\label{eq:Phalo}
	P_{\rm halo}(\mathrm{DM_{halo}^{MW}})=\frac{1}{ \sqrt{2 \pi} \sigma_{\mathrm {halo}}} \exp \left[-\frac{\left(\mathrm{DM_{halo}^{MW}}-\mu_{\mathrm {halo}}\right)^2}{2 \sigma_{\mathrm {halo}}^{2}}\right]\;.
\end{equation}

Based on the state-of-the-art IllustrisTNG simulation \citep{Springel2018}, \cite{Zhang2020} selected
a large sample of simulated galaxies with similar properties to observed FRB hosts to derive the
distributions of $\mathrm{DM}_{\mathrm {host}}$ of repeating and non-repeating FRBs. The distributions
of $\mathrm{DM}_{\mathrm {host}}$ can be well described by the log-normal function
\citep{Macquart2020,Zhang2020}
\begin{equation}\label{eq:Phost}
	P_{\rm host}\left(\mathrm{DM}_{\mathrm {host }} \right)=\frac{1}{\sqrt{2 \pi}\mathrm{DM}_{\mathrm {host }} \sigma_{\mathrm {host }} } \exp \left[-\frac{\left(\ln \mathrm{DM}_{\mathrm {host }}-\mu_{\mathrm {host}}\right)^2}{2 \sigma_{\mathrm {host}}^{2}}\right]\;,
\end{equation}
where $e^{\mu_{\mathrm{host}}}$ and
$e^{2\mu_{\mathrm{host}} + \sigma_{\mathrm{host}}^{2}}(e^{ \sigma_{\mathrm{host}}^{2}}-1)$
are the mean and variance of the distribution, respectively. Due to the diversity of host galaxies,
\cite{Zhang2020} computed the $\mathrm{DM_{host}}$ distributions for repeating FRBs in dwarf galaxies
(like FRB 121102, FRB 180301, FRB 181030, and FRB 190711), repeating FRBs in spiral galaxies
(like FRB 180916 and FRB 201124), and non-repeating FRBs separately. Here we divide the localized FRBs
into these three types according to their host properties. The evolution of the median of
$\mathrm{DM_{host}}$ ($\mathrm{\mu_{host}}$) can be fitted by $e^{\mathrm{\mu_{host}}}(z)=\kappa(1+z)^\gamma$, where
$\kappa$ and $\gamma$ are given by \cite{Zhang2020}. The propagated uncertainty $\sigma_{\mathrm {host}}$
of $\mathrm{DM_{host}}$ is calculated from the uncertainties of $\kappa$ and $\gamma$.
With this expression of redshift evolution, we can derive the $\mathrm{DM}_{\mathrm {host}}$ distributions
at any redshift of a localized FRB.

\begin{figure}
\vskip-0.1in
\centerline{\includegraphics[keepaspectratio,clip,width=0.5\textwidth]{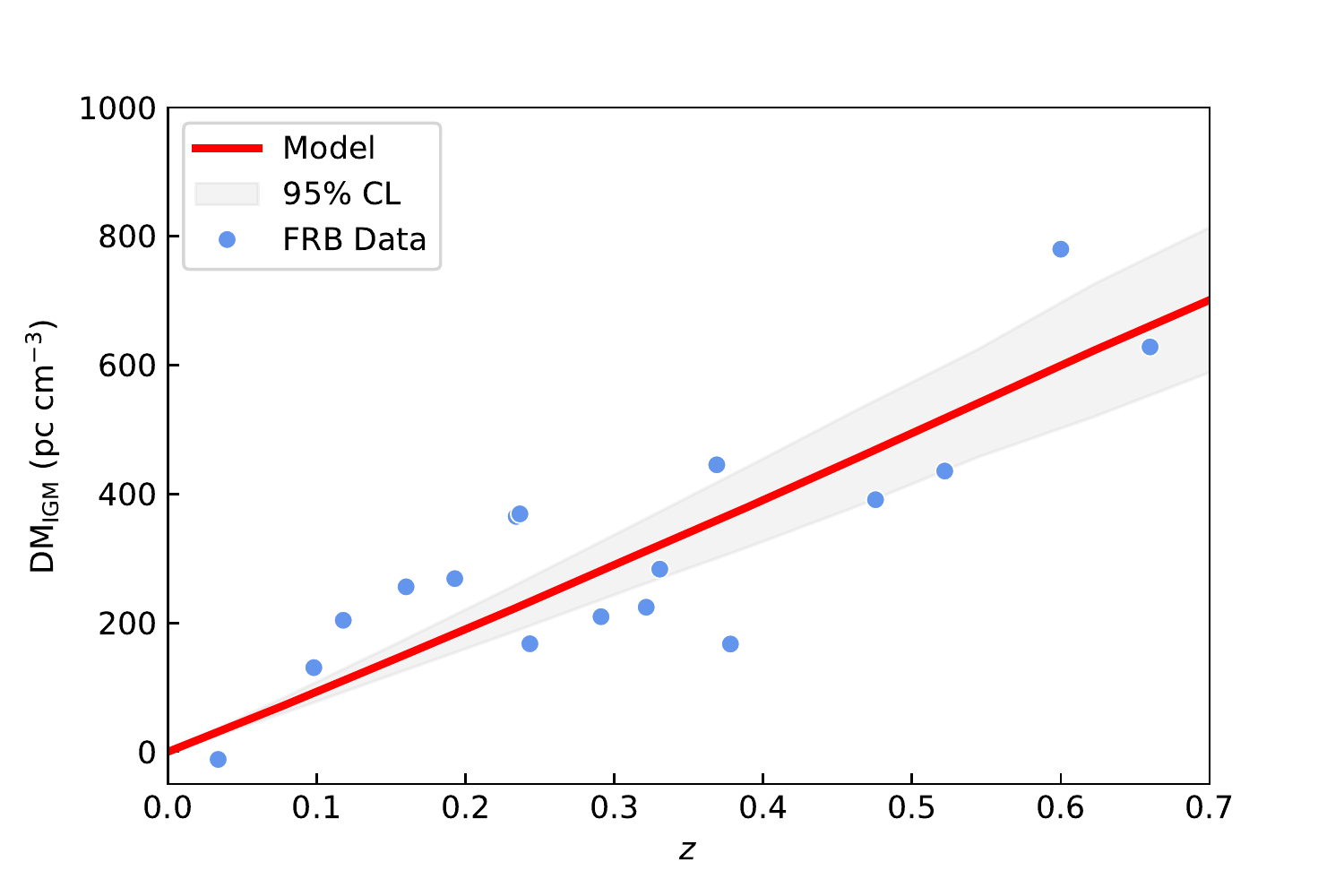}}
\caption{The $\mathrm{DM_{IGM}}$-$z$ relation for 17 localized FRBs.
The data points are estimations of $\mathrm{DM_{IGM}}$ versus redshift measurements
for the 17 localized FRBs. The $\mathrm{DM_{IGM}}$ values are estimated by correcting
the observed $\mathrm{DM_{obs}}$ for the contributions from our galaxy and the FRB host galaxy
(see the text for details). The solid line shows the model of Equation~(\ref{DMIGM}) with
the inferred parameters of $\Omega_{m}=0.309\pm0.006$, $\Omega_{b}h^2=0.02245\pm0.00013$,
$H_{0}=67.73\pm0.44$ $\mathrm{km\;s^{-1}\;Mpc^{-1}}$, and $f_{\mathrm{IGM,0}}=0.927\pm0.075$.
The shaded area represents the uncertainty of the estimated average $\mathrm{DM_{IGM}}$
value at the 95\% confidence level due to the uncertainties of the inferred model parameters which we mentioned above.}
\label{Fig1}
\end{figure}

To date, more than 20 FRBs have already been localized. Nonetheless, some of them are not
available for our analysis. For example, the DM of FRB 181030 is only 103.396 $\mathrm{pc\;cm^{-3}}$
\citep{Bhardwaj2021}, which will be reduced to a negative value after subtracting DM contributions
from the Milky Way ISM and halo. That is, the integral upper limit
($\mathrm{DM_{E}}\equiv\mathrm{DM_{obs}}-\mathrm{DM_{ISM}^{MW}}-\mathrm{DM_{halo}^{MW}}$)
in the probability of the external DM contribution outside our Galaxy (see Equation~(\ref{eq:P_FRB}))
will become negative. FRB 190520B is co-located with a compact, persistent radio source and associated
with a dwarf host galaxy at a redshift of 0.241 \citep{Niu2022}. It is a clear outlier from the
general trend of the extragalactic $\mathrm{DM_{E}}$--$z$ relation, with an unprecedented
DM contribution from its host galaxy. Thus its $\mathrm{DM_{host}}$ term can not be accurately deducted.
FRB 200110E is located in a globular cluster in the direction of the nearby galaxy M81
\citep{Bhardwaj2021b, Kirsten2022}. The distance of FRB 200110E is only 3.6 Mpc, and the IGM between
the Milky Way and M81 contributes of the order of $\mathrm{DM_{IGM}}\approx1$ $\mathrm{pc\;cm^{-3}}$.
Thus the cosmological information carried by FRB 200110E is too little. Additionally, the peculiar
velocity effect is significant, which makes it can not be used for cosmological studies. After excluding
these FRBs, we use a sample of 17 FRBs in the redshift range $0.0337\leq z \leq0.66$ to constrain
$f_{\mathrm{IGM}}(z)$.\footnote{After this work was done, we noticed a new article \citep{2022arXiv221004680R},
which has reported the discovery of a burst, FRB 220610A, in a complex host galaxy system at a redshift
of $z=1.016$. Whereas this burst was not included in our sample.} \autoref{Tab1} lists the redshifts,
$\mathrm{DM_{obs}}$, and $\mathrm{DM_{ISM}^{MW}}$ of our sample. The estimated $\mathrm{DM_{IGM}}$ and
measured $z$ values for the 17 localized FRBs are shown in Figure~\ref{Fig1}. We have estimated $\mathrm{DM_{IGM}}$
by subtracting the following from the observed $\mathrm{DM_{obs}}$ value: $\mathrm{DM_{ISM}^{MW}}$ from
the Galactic ISM model; a median of 65 $\mathrm{pc\;cm^{-3}}$ contributed by $\mathrm{DM_{halo}^{MW}}$;
and a median value of $\mathrm{DM_{host}}$ at different redshifts estimated from the IllustrisTNG simulation \citep{Zhang2020}.

\begin{table}
\centering \caption{Properties of the 17 localized FRBs}
\begin{tabular}{lcccc}
\hline
\hline
Name & Redshift & $\mathrm{DM_{obs}}$  & $\mathrm{DM_{ISM}^{MW}}$ & Refs.\\
     &          & $(\mathrm{pc\;cm^{-3}})$  & $(\mathrm{pc\;cm^{-3}})$ &          \\
\hline
FRB 121102 & 0.19273 & $557 \pm 2$ & 188.0 & 1  \\
FRB 180301 & 0.3304 & $536 \pm 0.2$ & 152.0 & 2   \\
FRB 180916 & 0.0337 & $349.349 \pm 0.005$ & 200.0 & 3 \\
FRB 180924 & 0.3214 & $361.42 \pm 0.06$ & 40.5 & 4 \\
FRB 181112 & 0.4755 & $589.27 \pm 0.03$ & 102.0 & 5 \\
FRB 190102 & 0.291 & $363.6 \pm 0.3$ & 57.3 & 6 \\
FRB 190523 & 0.66 & $760.8 \pm 0.6$ & 37.0 & 7  \\
FRB 190608 &  0.1178 & $338.7 \pm 0.5$ & 37.2 & 8  \\
FRB 190611 & 0.378 & $321.4 \pm 0.2$ & 57.83 & 9 \\
FRB 190614 & 0.6 & $959.2 \pm 0.5$ & 83.5 & 10  \\
FRB 190711 & 0.522 & $593.1 \pm 0.4$ & 56.4 & 9 \\
FRB 190714 & 0.2365 & $504 \pm 2$ & 38.0 & 9 \\
FRB 191001 & 0.234 & $506.92 \pm 0.04$ & 44.7 & 9  \\
FRB 191228 & 0.2432 & $297.5 \pm 0.05$ & 33.0 & 2  \\
FRB 200430 & 0.16 & $380.1 \pm 0.4$ & 27.0 & 9 \\
FRB 200906 & 0.3688 & $577.8 \pm 0.02$ & 36.0 & 2 \\
FRB 201124 & 0.098 & $413.52 \pm 0.05$ & 123.2 & 11 \\
\hline
\end{tabular}
\tablecomments{References: (1) \cite{Chatterjee2017}; (2) \cite{Bhandari2022};
(3) \cite{Marcote2020}; (4) \cite{Bannister2019}; (5) \cite{Prochaska2019b};
(6) \cite{Bhandari2020}; (7) \cite{Ravi2019}; (8) \cite{Chittidi2021};
(9) \cite{Heintz2020}; (10) \cite{Law2020}; (11) \cite{Ravi2022}.}
\label{Tab1}
\end{table}

To construct a likelihood function $\mathcal{L}$ from FRB measurements, we build a model for
$\mathrm{DM_{IGM}}$. The model probability distribution for $\mathrm{DM_{IGM}}$ has been derived
from the theoretical treatments of the IGM and galaxy halos with a standard deviation
$\sigma_{\mathrm{DM}}$ dominated by the variance in $\mathrm{DM_{IGM}}$. The $\mathrm{DM_{IGM}}$
distributions derived in both semi-analytic models and cosmological simulations can be well fitted
by a quasi-Gaussian function with a long tail \citep{McQuinn2014,Prochaska2019a,Macquart2020}
\begin{equation}\label{eq:PIGM}
	P_{\mathrm{IGM}}(\Delta)=A \Delta^{-\beta} \exp \left[-\frac{\left(\Delta^{-\epsilon}-C_{0}\right)^2}{2 \epsilon^{2} \sigma_{\mathrm{DM}}^{2}}\right],\;\;\; \Delta>0\;,
\end{equation}
where $\Delta\equiv\mathrm{DM}_{\mathrm{IGM}}/\left\langle\mathrm{DM}_{\mathrm{IGM}}\right\rangle$,
$A$ is a normalization coefficient, and the indices $\epsilon$ and $\beta$ are related to the inner
density profile of gas in halos. Here we take $\epsilon=3$ and $\beta=3$, as \cite{Macquart2020} did
in their treatment. $C_0$ is a free parameter, which can be fitted when the mean $\langle \Delta \rangle=1$.
The motivation for this analytic form (Equation~(\ref{eq:PIGM})) is that in the limit of small
$\sigma_{\mathrm{DM}}$, the $\mathrm{DM_{IGM}}$ distribution should approach a Gaussian owing to
the more diffuse halo gas and the Gaussianity of structure on large scales. Conversely, when the
variance is large, this probability distribution captures the large skew that due to a few large
structures that contribute to the DM of many sightlines. Recently, \cite{Zhang2021} used the
IllustrisTNG simulation to estimate the probability distributions of $\mathrm{DM_{IGM}}$
at different redshifts realistically. Following \cite{Wu2022}, the best-fit parameters ($A$, $C_0$, and
$\sigma_{\mathrm{DM}}$) of the $\mathrm{DM_{IGM}}$ distributions at the different redshifts that
presented by \cite{Zhang2021} are used for our purpose. The uncertainties of these
best-fit parameters may impact our final $f_{\rm IGM}$ constraint. To investigate whether
the uncertainties of these parameters affect the $\Omega_b$ constraint (similar to our $f_{\rm IGM}$
constraint), \cite{Yang2022} derived $\Omega_b$ using the best-fit values of these parameters
plus or minus the uncertainties. They found that these uncertainties have almost
no effect on the final $\Omega_b$ constraint. Given the fact that our $f_{\rm IGM}$ constraint
is almost the same as that of \cite{Yang2022}, we can come to the same conclusion.
Note that since the $\mathrm{DM_{IGM}}$ distributions are given in discrete redshifts \citep{Zhang2021},
we extrapolate them to the redshifts of the localized FRBs through cubic spline interpolation.

Given the model for $\mathrm{DM_{IGM}}$, we estimate the likelihood function by computing the joint
likelihoods of 17 FRBs \citep{Macquart2020}:
\begin{equation}
	\mathcal{L}_{\rm FRB}=\prod_{i=1}^{N_{\mathrm{FRB}}} P_{i}\left(\mathrm{DM}_{\mathrm{E}, i}\right)\;,
\end{equation}
where $P_{i}(\mathrm{DM}_{\mathrm{E}, i})$ is the probability of the total observed $\mathrm{DM_{obs}}$
corrected for our galaxy, i.e., $\mathrm{DM_{E}}\equiv\mathrm{DM_{obs}}-\mathrm{DM_{ISM}^{MW}}
-\mathrm{DM_{halo}^{MW}}=\mathrm{DM_{IGM}}+\mathrm{DM_{host}}/(1+z)$. For a burst at redshift $z_i$,
we have
\begin{eqnarray}
	P_{i}\left(\mathrm{DM}_{\mathrm{E}, i} \right) &=&
	\int_{0}^{\mathrm{DM}_{\mathrm{E}, i}}P_{\text {host}}\left(\mathrm{DM}_{\mathrm {host}}\right) \nonumber \\
	&\times&
	P_{\mathrm {IGM}}\left(\mathrm{DM}_{\mathrm{E}, i}-\mathrm{DM}_{\mathrm {host},z_i}\right) \mathrm{dDM}_{\mathrm {host}}\;,
\label{eq:P_FRB}
\end{eqnarray}
where the probability density functions for $P_{\text {host}}(\mathrm{DM_{host}})$ and
$P_{\text {IGM}}(\mathrm{DM_{IGM}})$ are obtained from Equations~(\ref{eq:Phost}) and (\ref{eq:PIGM}),
respectively. Note that the Milky Way halo DM distribution ($\mathrm{DM_{halo}^{MW}}$) will be considered as
a free parameter in our analysis. We will marginalize $\mathrm{DM_{halo}^{MW}}$ using a Gaussian prior of
$\mathrm{DM_{halo}^{MW}}=65\pm15$ $\mathrm{pc\;cm^{-3}}$ over the range of
$[\mu_{\mathrm {halo}}-3 \sigma_{\mathrm {halo}},\;\mu_{\mathrm {halo}}+ 3\sigma_{\mathrm {halo}}]$,
where $\mu_{\mathrm {halo}}=65$ $\mathrm{pc\;cm^{-3}}$ and $\sigma_{\mathrm {halo}}=15$ $\mathrm{pc\;cm^{-3}}$
(see Equation~(\ref{eq:Phalo})).

\section{Other Cosmological Probes}
\label{Sec:probes}
As we discussed in Section~\ref{Sec:Intro}, in order to break the degeneracy between the IGM
baryon fraction $f_{\mathrm{IGM}}$ and other cosmological parameters ($\Omega_{m}$, $\Omega_{b}$,
and $H_{0}$; see Equation~(\ref{DMIGM})), we use up-to-date cosmological data compilations,
including SNe Ia, BAO, and CMB data. The exact compilations and the details for
the likelihoods are described separately in what follows.

\subsection{Type Ia Supernovae}
The SNe Ia dataset that we use in this work is the Pantheon sample, which consists of 1048
SNe Ia in the redshift range $0.01 < z < 2.3$ \citep{Scolnic2018}. The observed distance modulus
of each SN is given as
\begin{equation}
	\mu_{\rm SN} = m_{\rm corr}-M_{B}\;,
\end{equation}
where $m_{\rm corr}$ is the corrected apparent magnitude and $M_{B}$ is the absolute magnitude.

The theoretical distance modulus $\mu_{\rm th}(z)$ is defined as
\begin{equation}
	\mu_{\rm th}= 5\log_{10} \left [ \frac{d_{L}(z)}{\mathrm{Mpc}}  \right ] +25\;,
\end{equation}
where $d_L(z)=(1+z)\frac{c}{H_0}\int_{0}^{z}{\frac{dz'}{\sqrt{\Omega_{m}(1+z')^3+1-\Omega_{m}}}}$
is the luminosity distance in the flat $\Lambda$CDM model. Thus, the $\chi^2_{\rm SN}$ function
for the Pantheon data is
\begin{equation}
	\chi^2_{\rm SN}=\Delta \bf{\hat{\mu}}^{\emph{T}} \cdot C_{\rm SN}^{-1} \cdot \Delta \bf{\hat{\mu}}\;,
\end{equation}
where $\Delta \hat{\mu}=\hat{\mu}_{\rm SN}-\hat{\mu}_{\rm th}$ is the data vector, defined by
the difference between the SN distance modulus $\mu_{\rm SN}$ and the theoretical distance modulus
$\mu_{\rm th}$, and $\bf{C_{\rm SN}}$ is the covariance matrix that contains both statistical
and systematic uncertainties of SNe.

\subsection{Baryon Acoustic Oscillations}
Primordial perturbations in the early universe excite acoustic waves in the plasma, known as BAO.
After the recombination period, the propagation of acoustic waves was frozen. Thus, there is a
characteristic scale called the comoving sound horizon $r_{s}$, which can be approximated as \citep{Aubourg2015}
\begin{eqnarray}
	r_{s}(z^\ast)&=&\int_{z^{\ast}}^{\infty} \frac{c_{s}(z)}{H(z)} dz  \nonumber \\
	&\approx& \frac{55.154 \exp \left[-72.3\left(\Omega_{v} h^{2}+0.0006\right)^{2}\right]}{\left(\Omega_{b} h^{2}\right)^{0.12807}\left(\Omega_{m} h^{2}\right)^{0.25351}}\; \mathrm{Mpc}\;,
\end{eqnarray}
where $c_s(z)$ is the sound speed of the photon-baryon fluid, $z$ is the redshift at the drag epoch,
$\Omega_{v}$ is the present-day neutrino density, and $h\equiv H_0/(100\;
\mathrm{km\; s^{-1}\; Mpc^{-1}})$ is the reduced Hubble constant. Here we use a combination of 11 BAO measurements
from \cite{Ryan2019}. Six of these BAO measurements are correlated, in which case $\chi^2$ is given by
\begin{equation}
	 \chi^2_{\rm BAO1}=\left(\bf{A_{obs}}-\bf{A_{th}}\right)^{\emph{T}} \cdot \bf{C_{\rm BAO}^{-1}} \cdot \left(\bf{A_{obs}}-\bf{A_{th}}\right)\;,
\end{equation}
where $\bf{A_{obs}}$ ($\bf{A_{th}}$) is the vector that contains all of the six measured (theoretical) values
and $\bf{C_{\rm BAO}}$ is the covariance matrix for the BAO data sets. The other five BAO measurements are
uncorrelated, so
\begin{equation}
	\chi^2_{\rm BAO2}=\sum_{i=1}^{5} \frac{\left[A_{\rm th}(z_i)-A_{\rm obs}(z_i)\right]^{2}}{\sigma_{i}^{2}}\;,
\end{equation}	
where $\sigma_{i}$ is the standard deviation of the $i$-th BAO measurement $A_{\rm obs}(z_i)$. The data are
combined into a $\chi^2$-statistic $\chi^2_{\rm BAO}=\chi^2_{\rm BAO1}+\chi^2_{\rm BAO2}$.

\subsection{Cosmic Microwave Background}
For the CMB measurements, we use the derived parameters, including the acoustic scale $l_A$, the shift parameter $R$,
and $\Omega_{b}h^2$ from the Planck analysis of the CMB (TT, TE, EE $+$ lowE) \citep{Chen2019,2020A&A...641A...6P}.
The acoustic scale is
\begin{equation}\label{eq:lA}
	l_A(z^\ast)=(1+z^\ast)\frac{\pi d_A(z^\ast)}{r_s(z^\ast)}\;,
\end{equation}
where $r_s$ is the comoving sound horizon at the recombination and $d_A=d_L(1+z)^{-2}$ is the angular diameter
distance. The shift parameter is
\begin{equation}\label{eq:R}
	R(z^\ast)=(1+z^\ast)\frac{d_A(z^\ast) \sqrt{\Omega_m}H_0}{c}\;.
\end{equation}
The redshift at decoupling $z^\ast$ is given by
\begin{equation}
	z^{\ast}=1048\left[1+0.00124\left(\Omega_{b} h^{2}\right)^{-0.738}\right]\left[1+g_{1}\left(\Omega_{m} h^{2}\right)^{g_{2}}\right]\;,
\end{equation}	
\begin{equation}
	g_{1} =\frac{0.0738\left(\Omega_{b} h^{2}\right)^{-0.238}}{1+39.5\left(\Omega_{b} h^{2}\right)^{0.763}}\;,
\end{equation}	
\begin{equation}
	g_{2} =\frac{0.560}{1+21.1\left(\Omega_{b} h^{2}\right)^{1.81}}\;.
\end{equation}

By setting $\boldsymbol{x}=(R,\;l_A,\;\Omega_{b}h^2)$, the $\chi^2_{\rm CMB}$ value for the CMB data is then
\begin{equation}
	 \chi^2_{\rm CMB}=\left(\boldsymbol{x}_{\rm obs}-\boldsymbol{x}_{\rm th}\right)^{\emph{T}} \cdot  \bf{C_{\rm CMB}^{-1}} \cdot \left(\boldsymbol{x}_{\rm obs}-\boldsymbol{x}_{\rm th}\right),
\end{equation}
where $\boldsymbol{x}_{\rm obs}=(1.7502,\;301.471,\;0.02236)$ and $\boldsymbol{x}_{\rm th}$ contain the observed and
theoretical values of the derived parameters, respectively, and $\bf{C_{\rm CMB}}$ is the covariance matrix
for the CMB data. Note that the distance priors derived from the CMB data are dependent on the specific cosmological model.
Here we adopt the values of $\boldsymbol{x}_{\rm obs}$ and $\bf{C_{\rm CMB}}$ that inferred from the flat $\Lambda$CDM model.

\section{Parameter Estimate and Results}
\label{Sec:Results}
To assess how well localized FRBs may help to constrain the evolution of the IGM baryon fraction
$f_{\mathrm{IGM}}(z)$, we consider two different parametric models. First, a simple constant model,
\begin{equation}
	f_{\rm{IGM}} = f_{\rm{IGM},0}\;.
\label{eq:constant}
\end{equation}
And second, a time-dependent model given by
\begin{equation}
	f_{\mathrm{IGM}}(z)=f_{\mathrm{IGM},0}\left(1+\alpha \frac{z}{1+z}\right)\;,
\label{eq:time-dependent}
\end{equation}
where $f_{\rm{IGM},0}$ is the present value of $f_{\rm{IGM}}$ and $\alpha$ quantifies any possible evolution
of $f_{\rm{IGM}}$. As massive halos are more abundant in the late universe, $f_{\rm{IGM}}$ is believed to
grow with redshift \citep{McQuinn2014,Prochaska2019a}. Therefore, here we require $\alpha\ge0$.

The quantities $f_{\mathrm{IGM}}(z)$, $\Omega_{m}$, $\Omega_{b}h^2$, $H_0$, $M_B$, and $\mathrm{DM_{halo}^{MW}}$
are fitted to the FRB, SN Ia, BAO, and CMB data simultaneously using the Python MCMC module $emcee$
\citep{Foreman2013}. Given the relation $\mathcal{L}\propto \exp[-\chi^2/2]$, the final log-likelihood
sampled by $emcee$ is a sum of the separated likelihoods of FRBs, SNe Ia, BAO, and CMB:
\begin{equation}
	\ln \mathcal{L}_{\rm tot}=\ln \mathcal{L}_{\rm FRB}-\chi^2_{\rm SN}/2-\chi^2_{\rm BAO}/2-\chi^2_{\rm CMB}/2\;.
\end{equation}
In our baseline analysis, we set flat priors on $f_{\mathrm{IGM,0}}\in[0,\;1]$ and $\alpha\in[0,\;2]$,
and a Gaussian prior on $\mathrm{DM_{halo}^{MW}}=65\pm15$ $\mathrm{pc\;cm^{-3}}$ over the 3$\sigma$ range
of $[20,\;110]$ $\mathrm{pc\;cm^{-3}}$.

\begin{figure*}
	\centering
	\includegraphics[width=0.75\textwidth]{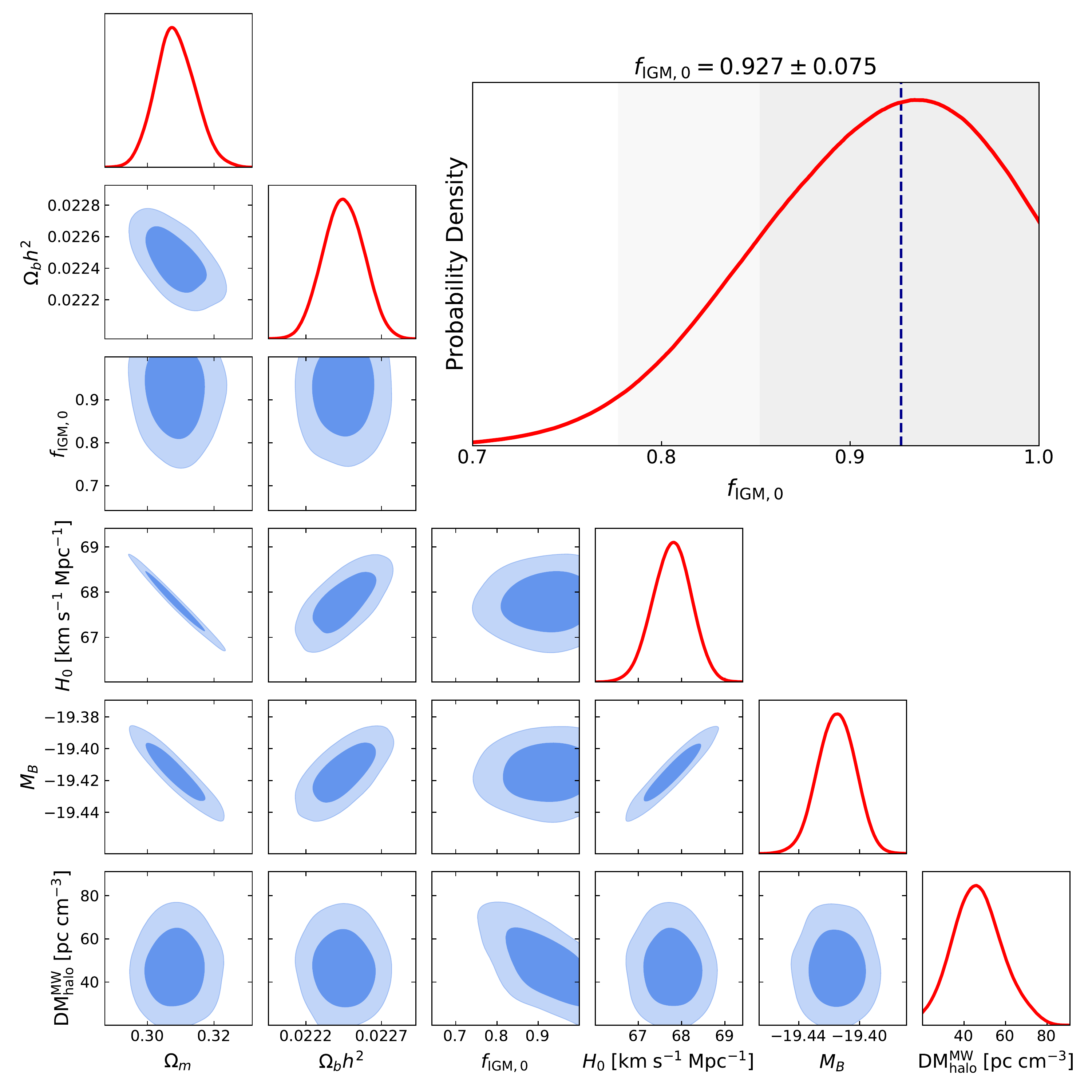}
	\caption{1D and 2D marginalized posterior distributions with the $1-2\sigma$ contours for the parameters $f_{\mathrm{IGM,0}}$, $\Omega_{m}$, $\Omega_{b}h^2$, $H_0$, $M_B$, and $\mathrm{DM_{halo}^{MW}}$ in the constant case of $f_{\rm{IGM}} = f_{\rm{IGM},0}$, constrained by the FRB, SN Ia, BAO, and CMB data. The 1D marginalized posterior distribution of $f_{\mathrm{IGM,0}}$ is magnified
	on the top-right panel. The vertical dashed line represents the best fit, whereas the shaded areas correspond to the 68\% and 95\% confidence regions.}
	\label{fig2}
\end{figure*}

\begin{deluxetable}{lcc}
	\centering
	\tablecaption{Constraints on all parameters for two different parametric models of $f_{\mathrm{IGM}}$. \label{table2}}
	\tablewidth{0pt}
	\tablehead{
					Model & Constant & Time-dependent\\
     \hline
                    Parameter & \multicolumn{2}{c}{Estimation with 68\% limits}
	}
	\startdata
					\hline
					$ f_{\mathrm{IGM,0}}$ & $0.927\pm0.075$ &  $0.837\pm0.089$ \\
					$\alpha$ & -- &  $<0.882$  \\
					$\Omega_{m}$ & $0.309\pm0.006$  & $0.309\pm0.006$ \\
					$\Omega_{b}h^2$ & $0.02245\pm0.00013$  & $0.02245\pm0.00013$ \\
					$H_{0}/[\mathrm{km\; s^{-1}\; Mpc^{-1}}]$ & $67.78\pm0.44$  &  $67.71\pm0.42$\\
					$M_B$ & $-19.415\pm0.012$ &  $-19.417\pm0.012$ \\
					$\mathrm{DM_{halo}^{MW}}/[\mathrm{pc\;cm^{-3}}]$ & $47\pm10$ &   $49\pm10$  \\
					\hline
					$-2 \ln \mathcal{L}_{\rm max}$ & $1262.855$  & $1263.099$ \\
					$\Delta$AIC & --  &  $2.244$ \\
	\enddata
\end{deluxetable}

For the constant case, there are six free parameters, including the IGM baryon fraction $f_{\mathrm{IGM,0}}$,
the cosmological parameters ($\Omega_{m}$, $\Omega_{b}h^2$, and $H_0$), the SN absolute magnitude $M_B$, and the Milky Way halo DM contribution $\mathrm{DM_{halo}^{MW}}$. The 1D marginalized posterior distributions and 2D plots of the $1-2\sigma$ confidence regions for these six parameters are displayed in Figure~\ref{fig2}.
These contours show that, at the $1\sigma$ confidence level, the inferred parameter values are $f_{\mathrm{IGM,0}}=0.927\pm0.075$, $\Omega_{m}=0.309\pm0.006$, $\Omega_{b}h^2=0.02245\pm0.00013$, $H_0=67.78\pm0.44$ $\mathrm{km\;s^{-1}\;Mpc^{-1}}$, and $\mathrm{DM_{halo}^{MW}}=47\pm10$ $\mathrm{pc\;cm^{-3}}$.
The corresponding results are summarized in Table~\ref{table2}. Figure~\ref{Fig1} also shows the theoretical curve for $\langle\mathrm{DM_{IGM}}\rangle$ versus $z$ for the constant case and a model estimate of the scatter (95\% interval) due to the uncertainties of the inferred parameters. The theoretical curve reflects the trend of the data well.
Moreover, the inferred value of the IGM baryon fraction is compatible with previous results obtained from observations (e.g., \citealt{1998ApJ...503..518F,2004ApJ...616..643F,Shull2012,2016PhRvL.117e1301H, 2018PhRvD..98j3518M}) and simulations (e.g., \citealt{1999ApJ...514....1C,2006ApJ...650..560C}). The constraint accuracy of $f_{\mathrm{IGM,0}}$ is about 8.0\%.

\begin{figure*}
	\centering
	\includegraphics[width=0.85\textwidth]{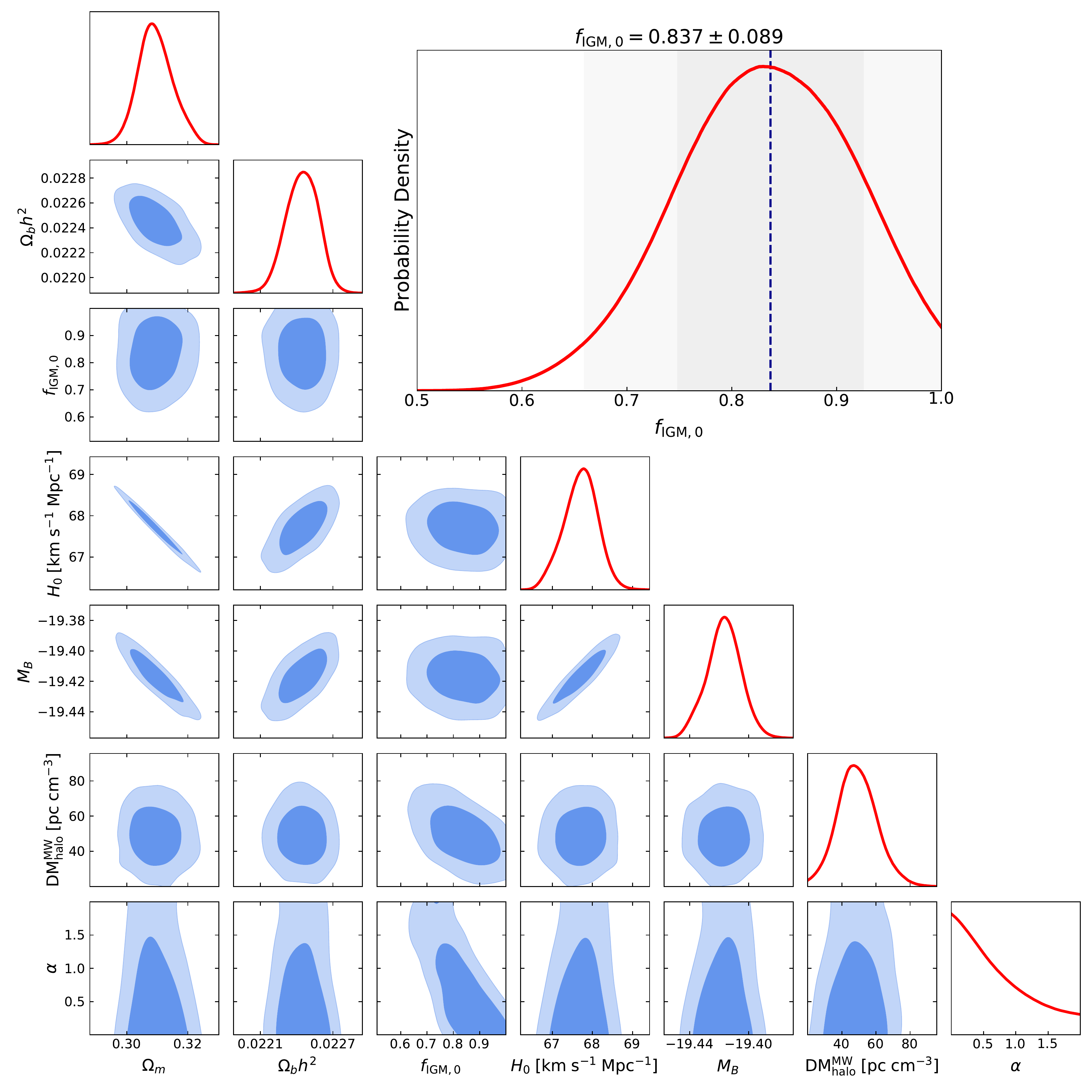}
	\caption{Same as Figure~\ref{fig2}, but now for the time-dependent case of $f_{\mathrm{IGM}}(z)=f_{\mathrm{IGM},0}(1+\alpha \frac{z}{1+z})$.}
	\label{fig3}
\end{figure*}

For the time-dependent case, the free parameters are $\{f_{\mathrm{IGM,0}},\;\alpha,\;\Omega_{m},\;\Omega_{b}h^{2},\;H_0,\;M_B,\;\mathrm{DM_{halo}^{MW}}\}$.
These seven parameters are constrained to be $f_{\mathrm{IGM,0}}=0.837\pm0.089$, $\alpha<0.882$, $\Omega_{m}=0.309\pm0.006$, $\Omega_{b}h^2=0.02245\pm0.00013$, $H_0=67.71\pm0.42$ $\mathrm{km\;s^{-1}\;Mpc^{-1}}$, and $\mathrm{DM_{halo}^{MW}}=49\pm10$ $\mathrm{pc\;cm^{-3}}$, which are displayed in Figure~\ref{fig3} and summarized in Table~\ref{table2}.
Note that with the requirement that $f_{\mathrm{IGM}}$ grows with redshift ($\alpha\ge0$), only an upper limit on $\alpha$ can be estimated, which implies that there is no strong evidence for the redshift dependence of $f_{\mathrm{IGM}}$.
This is consistent with the cosmology-insensitive result obtained from five localized FRBs \citep{Li2020}. The comparison between columns 2 and 3 of Table~\ref{table2} suggests that the nuisance parameters ($\Omega_{m}$, $\Omega_{b}h^2$, $H_0$, $M_B$, and $\mathrm{DM_{halo}^{MW}}$) are almost identical and have little effect on the adopted parametric model of $f_{\mathrm{IGM}}$.

Because the constant model and the time-dependent model do not have the same number of free parameters, a comparison of the likelihoods for either being closer to the correct model must be based on model selection criteria.
We use the Akaike Information Criterion (AIC; \citealt{Akaike1974, Akaike1981}) to test the statistical performance of the models, $\mathrm{AIC} =  -2 \ln \mathcal{L}+2p$, where $p$ is the number of free parameters.
With $\mathrm{AIC}_1$ and $\mathrm{AIC}_2$ characterizing models $\mathcal{M}_1$ (the constant model) and $\mathcal{M}_2$ (the time-dependent model), respectively, the difference $\Delta \mathrm{AIC}\equiv\mathrm{AIC}_{2}-\mathrm{AIC}_{1}$ determines the extent to which $\mathcal{M}_1$ is favoured over $\mathcal{M}_2$.
The evidence of $\mathcal{M}_1$ being correct is judged `weak' when the outcome $\Delta \equiv\mathrm{AIC}_{2}-\mathrm{AIC}_{1}$ is in the range $0<\Delta<2$, `positive' when $2<\Delta<6$, and `strong' when $\Delta>6$.
Therefore, the outcome $\Delta \mathrm{AIC}=2.244$ shows a positive evidence in favor of the constant model ($\mathcal{M}_1$) with respect to the time-dependent model ($\mathcal{M}_2$).
Nevertheless, we hold the opinion that this positive evidence may be due to the relatively low redshifts in the FRB data.
To distinguish between the constant and time-dependent models better, a larger number of FRBs localized at higher redshifts is required in the future.

\begin{figure*}
\centering
\includegraphics[width=0.48\textwidth]{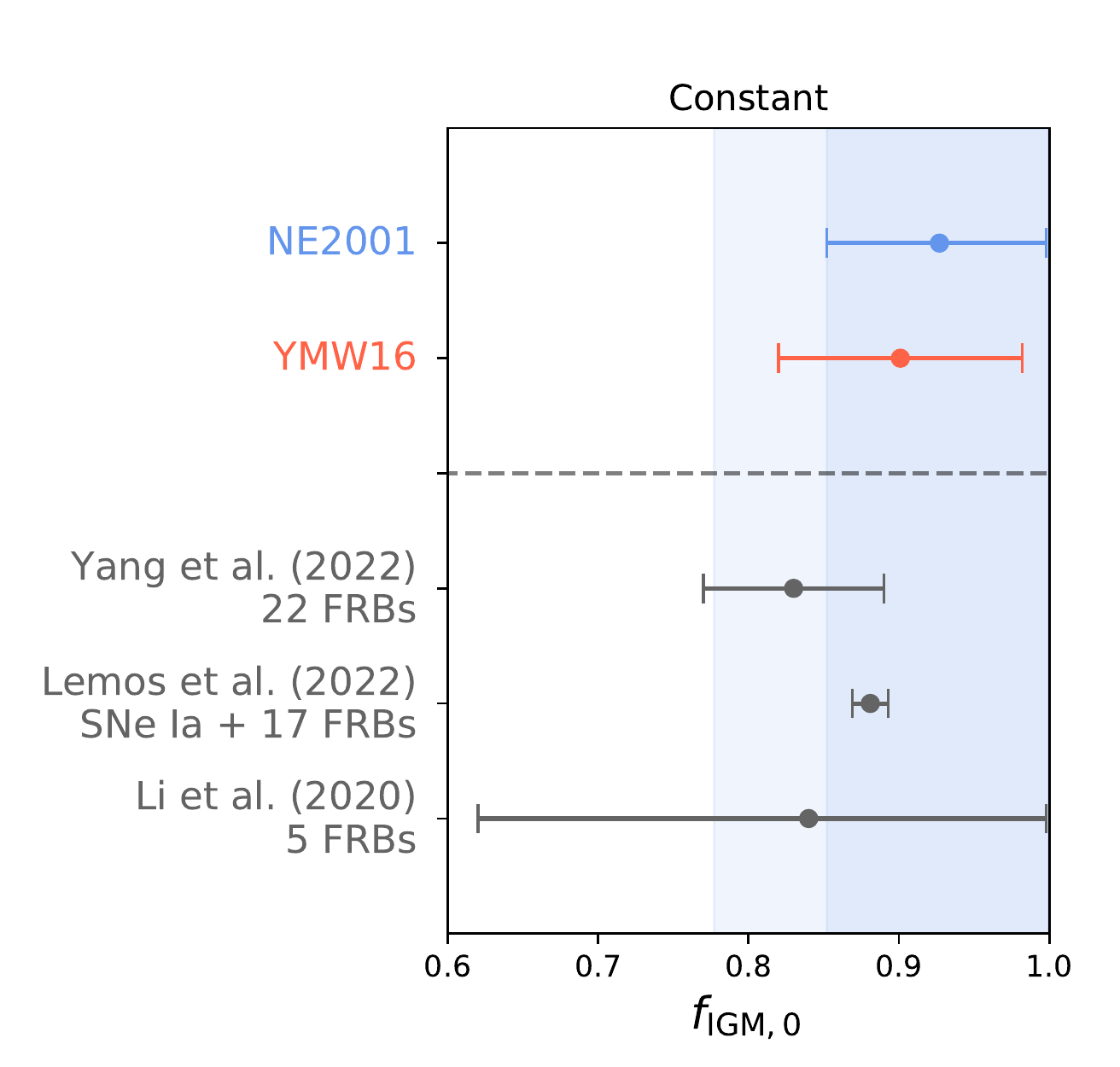}
\includegraphics[width=0.48\textwidth]{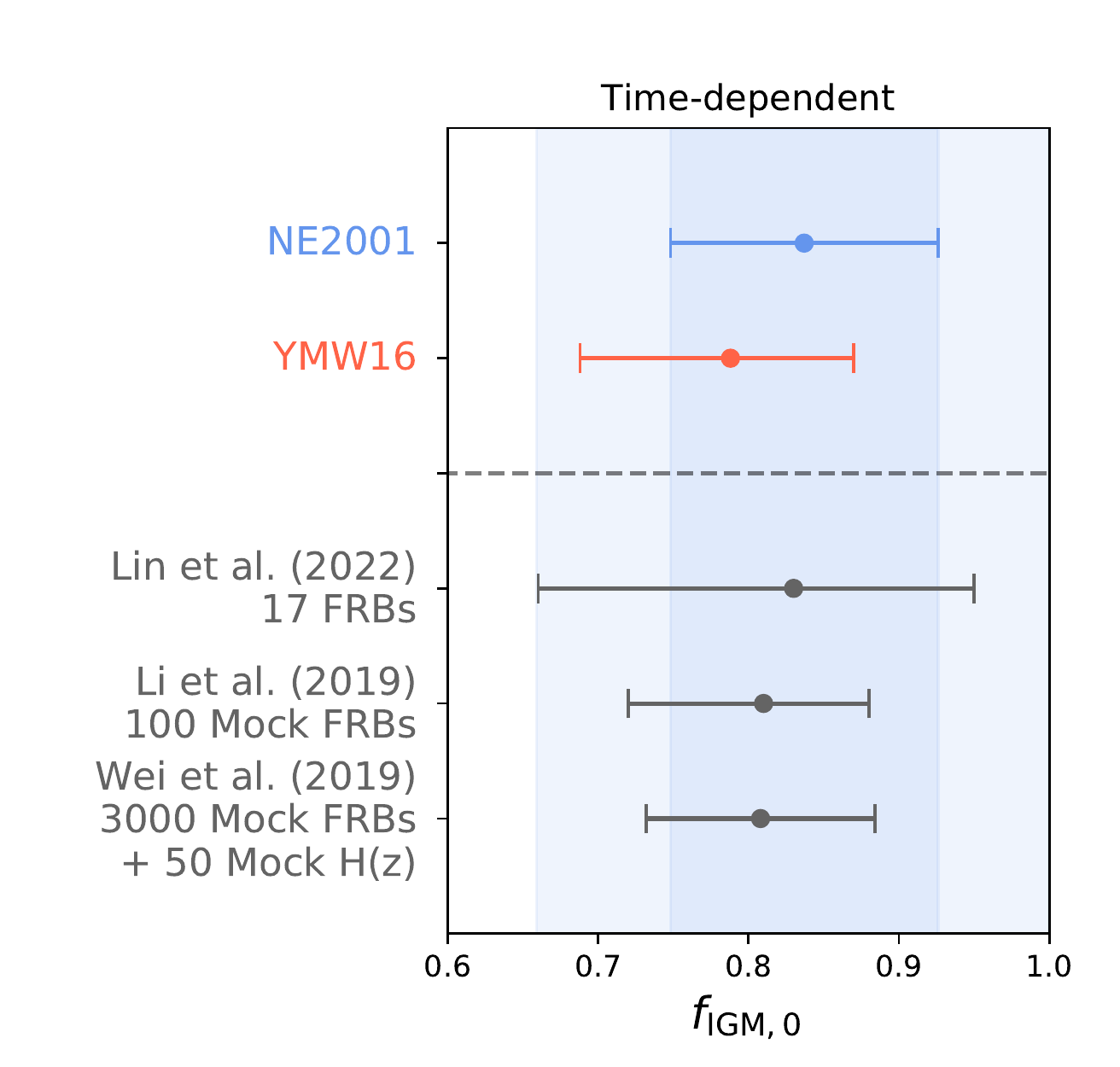}
\vskip-0.1in
\caption{Comparison of $f_{\mathrm{IGM,0}}$ constraints with previous results in the constant (left panel) and time-dependent (right panel) cases. The shaded areas in both panels represent the 1$\sigma$ and 2$\sigma$ confidence regions of the corresponding $f_{\mathrm{IGM,0}}$ constraints.}
\label{fig4}
\end{figure*}

In our analysis, the $\mathrm{DM_{ISM}^{MW}}$ values are estimated from the Galactic electron density model of
NE2001 \citep{Cordes2002}. We also perform a parallel comparative analysis of the FRB data using the YMW16 model \citep{Yao2017}.
The resulting constraints now turn to be $f_{\mathrm{IGM,0}}=0.901\pm 0.081$ ($f_{\mathrm{IGM,0}}=0.788^{+0.082}_{-0.100}$)
for the constant (time-dependent) case. Comparing these inferred $f_{\mathrm{IGM,0}}$ with those obtained using the NE2001 model,
we see that the adoption of a different electron distribution model has a minimal influence on the results.
Due to the larger $\mathrm{DM_{ISM}^{MW}}$ contribution at low Galactic latitudes in the YMW16 Model, the derived $f_{\mathrm{IGM,0}}$ values are slightly smaller than those in the NE2001 model.

To make a direct comparison with previous works, in Figure~\ref{fig4} we plot some typical $f_{\mathrm{IGM,0}}$ constraints
from different FRB samples, as well as our constraints from both the NE2001 and YMW16 models. One can see from Figure~\ref{fig4}
that our $f_{\mathrm{IGM,0}}$ constraints are well consistent with previous results at the $1\sigma$ confidence level.

\section{Conclusions}
\label{Sec:Conclusion}
The $\mathrm{DM_{IGM}}$-$z$ relation of FRBs has been used for probing the baryon fraction in the IGM,
$f_{\mathrm{IGM}}$. However, such studies have been restricted by the strong degeneracy between
cosmological parameters and $f_{\mathrm{IGM}}$. Moreover, the DM contribution from the IGM
($\mathrm{DM_{IGM}}$) cannot be effectively distinguished from other DMs contributed by the Milky Way
or host galaxy. In this work, we investigate precise constraints on $f_{\mathrm{IGM}}$ from
the DM measurements of seventeen localized FRBs. In order to break the parameter degeneracy,
we combine FRB data with three other cosmological probes (including SNe Ia, BAO, and CMB) to
infer cosmological parameters and $f_{\mathrm{IGM}}$ simultaneously. To avoid uncontrollable
systematic errors induced by the $\mathrm{DM_{IGM}}$ deduction, we handle the DM contributions
of the host galaxies and IGM as the probability distributions derived from the the IllustrisTNG simulation.

Following the analysis method described in Section~\ref{Sec:Dispersion}, we explore the possible
redshift dependence of $f_{\mathrm{IGM}}(z)$ considering two different parametric models, which
are expressed as the constant and time-dependent parameterizations given by Equations~(\ref{eq:constant})
and (\ref{eq:time-dependent}). The MCMC analysis is used to constrain $f_{\mathrm{IGM}}(z)$ and other
cosmological parameters. For the constant model, we infer that $f_{\mathrm{IGM,0}}=0.927\pm0.075$,
representing a precision of 8.0\%. This constraint from FRB observations is roughly consistent with
those obtained from other probes \citep{1998ApJ...503..518F,2004ApJ...616..643F,Shull2012,2016PhRvL.117e1301H}.
For the time-dependent model, whereas only an upper limit on the evolution index $\alpha$ can be set
($\alpha<0.882$), we can obtain a good limit on the local $f_{\mathrm{IGM,0}}=0.837\pm0.089$,
which is slightly looser but still consistent with previous results derived from different methods.
According to the AIC model selection criteria, there is a mild evidence suggesting that the constant
model is preferred over the time-dependent model. However, due to the fact that the number of current
localized FRBs is small and their redshift measurements are relatively low, we cannot safely exclude
the possibility of an evolving $f_{\mathrm{IGM}}(z)$.

Redshift measurements of a larger sample of FRBs are essential for using this method presented here
to constrain $f_{\mathrm{IGM}}$ and its possible redshift evolution. Forthcoming radio telescopes
such as the Deep Synoptic Array 2000-dish prototype \citep{2019BAAS...51g.255H} and the Square Kilometre
Array \citep{2009IEEEP..97.1482D}, with improved detection sensitivity and localization capability,
will be able to increase the current localized FRB sample size by orders of magnitude. With the
rapid progress in localizing FRBs, the constraints on $f_{\mathrm{IGM}}$ will be significantly improved,
and the baryon distribution of the universe will be better understood.

\begin{acknowledgments}
We are grateful to the anonymous referee for their helpful comments.
This work is partially supported by the National Key Research and Development Program of
China (2022SKA0130100), the National Natural Science Foundation of China (grant Nos.
11725314 and 12041306), the Key Research Program of Frontier Sciences (grant No. ZDBS-LY-7014)
of Chinese Academy of Sciences, International Partnership Program of Chinese Academy of Sciences
for Grand Challenges (114332KYSB20210018), the CAS Project for Young Scientists in Basic Research
(grant No. YSBR-063), the CAS Organizational Scientific Research Platform for National Major
Scientific and Technological Infrastructure: Cosmic Transients with FAST, the Natural Science
Foundation of Jiangsu Province (grant No. BK20221562), and the Young Elite Scientists
Sponsorship Program of Jiangsu Association for Science and Technology.
\end{acknowledgments}


\end{document}